\documentclass[twocolumn, superscriptaddress,showpacs,amsmath,amssymb,aps,pra,floatfix, 10pt]{revtex4-1}
\usepackage[latin1]{inputenc}
\usepackage[american]{babel}
\usepackage[T1]{fontenc}
\usepackage{mathrsfs}
\usepackage{hyperref}
\usepackage{multirow}
\usepackage{bm, epsfig}
\usepackage{graphicx}
\usepackage{subeqnarray}
\usepackage{bbold}
\usepackage{upgreek}
\usepackage[normalem]{ulem}

\hypersetup{colorlinks=true, urlcolor=blue, citecolor=blue, filecolor=blue, linkcolor=blue}
\newcommand*{\diff}{\mathop{}\!\mathrm{d}}

\newcommand{\uimm}{\mathrm{i}}
\newcommand{\eu}{\mathrm{e}}
\newcommand{\daga}{^{\dagger}}
\newcommand{\unit}[1]{\ensuremath{\, \mathrm{#1}}}
\newcommand*{\sinc}{\mathop{}\!\mathrm{sinc}}

\graphicspath{{Images/}}

\begin{document}

\title{\mbox{X-ray} frequency combs from optically controlled resonance fluorescence}

\date{\today}

\author{Stefano~M.~Cavaletto}
\email[Corresponding author: ]{smcavaletto@gmail.com}
\affiliation{Max-Planck-Institut f\"{u}r Kernphysik, Saupfercheckweg 1, 69117 Heidelberg, Germany}
\author{Zolt\'{a}n~Harman}
\affiliation{Max-Planck-Institut f\"{u}r Kernphysik, Saupfercheckweg 1, 69117 Heidelberg, Germany}
\affiliation{ExtreMe Matter Institute (EMMI), Planckstrasse 1, 64291 Darmstadt, Germany}
\author{Christian~Buth}
\altaffiliation[Present address: ]{Max-Planck-Institut f\"ur Quantenoptik, Hans-Kopfermann-Stra\ss{}e~1, 85748~Garching bei M\"unchen, Germany}
\affiliation{Max-Planck-Institut f\"{u}r Kernphysik, Saupfercheckweg 1, 69117 Heidelberg, Germany}
\author{Christoph~H.~Keitel}
\affiliation{Max-Planck-Institut f\"{u}r Kernphysik, Saupfercheckweg 1, 69117 Heidelberg, Germany}

\begin{abstract}
An \mbox{x-ray} pulse-shaping scheme is put forward for imprinting an optical frequency comb onto the radiation emitted on a driven \mbox{x-ray} transition, thus producing an \mbox{x-ray} frequency comb. A four-level system is used to describe the level structure of $N$ ions driven by narrow-bandwidth x~rays, an optical auxiliary laser, and an optical frequency comb. By including many-particle enhancement of the emitted resonance fluorescence, a spectrum is predicted consisting of {equally spaced narrow lines which are} centered on an \mbox{x-ray} transition energy and separated by the same tooth spacing as the driving optical frequency comb. {Given an \mbox{x-ray} reference frequency, our comb could be employed to determine an unknown \mbox{x-ray} frequency. While relying on the quality of the light fields used to drive the ensemble of ions, the model has validity at energies from the 100~eV to the keV range.}
\end{abstract}


\maketitle


\section{Introduction}

Major advances in metrology and precision spectroscopy were led by the introduction \cite{Udem:99, *Reichert199959} and development \cite{PhysRevLett.82.3568, *PhysRevLett.84.3232, *Jones28042000, *PhysRevLett.84.5102, *PhysRevLett.84.5496, *PhysRevLett.85.2264} of optical frequency combs \cite{Nature.416.233, *RevModPhys.75.325}.  The spectrum of a frequency comb consists of a series of equally spaced teeth, i.e., modes of a train of femtosecond pulses spaced by the repetition frequency of a mode-locked laser. By counting the number of teeth between an unknown optical frequency and an optical reference line, this comb is used as a fine ruler to measure an optical frequency instead of the corresponding wavelength, which can be determined much less precisely. This allows one to reach relative accuracies up to $10^{-18}$ \cite{1402-4896-86-6-068101}. By precisely counting optical oscillations, e.g., in trapped-atom and \mbox{-ion} standards, optical frequency combs play a crucial role in the realization of all-optical atomic clocks \cite{RevModPhys.78.1279, Diddams03082001}.

In light of the success of optical-frequency-comb metrology, it is desirable {as an ultimate aim} to render this technology available for extreme ultraviolet (XUV) and \mbox{x-ray} frequencies \cite{RevModPhys.78.1297}. \mbox{X-ray} frequency combs promise to enable precise measurements of high-energy transitions paralleling the accuracy achieved for optical frequencies, with an improvement of several orders of magnitude. This is anticipated to allow, to name but a few examples, even more stringent experimental tests of quantum electrodynamics and astrophysical models \cite{BernittNature}, and the search for the variability of the fine-structure constant, to which transitions in highly charged ions are predicted to be more sensitive \cite{PhysRevLett.106.210802}. One may also {eventually} envision ultraprecise \mbox{x-ray} atomic clocks.  

XUV frequency combs have been generated via intracavity high-order harmonic generation (HHG) \cite{PhysRevLett.94.193201, *Nature.436.234}. While in conventional HHG an optical pulse in a gas produces a spectrum of odd harmonics of the optical frequency, in intracavity HHG a train of coherent optical pulses generates a spectrum which in each harmonic line is structured into a fine comb. Based on this scheme, Ref.~\cite{Nature.482.68} reported the observation of frequency combs at wavelengths of $\sim$\,$40\,\unit{nm}$ (photon energy of $\sim$\,$30\,\unit{eV}$). The required optical peak intensity of $\sim$\,$10^{14}\,\unit{W/cm^2}$ was obtained with a femtosecond enhancement cavity. Yet relativistic effects limit the range in which HHG operates efficiently \cite{Kohler2012159}, {i.e., where \mbox{x-ray} frequency combs are presently advisable with HHG-based methods. Investigations of alternative schemes are, therefore, timely.}

Here, we put forward a scheme for coherent \mbox{x-ray} pulse shaping to imprint the structure of an optical frequency comb onto the resonance fluorescence spectrum that is emitted on an \mbox{x-ray} transition. We refer to previous investigations of many-color schemes of resonance fluorescence in multi-level systems \cite{PhysRevA.42.1630, *PhysRevA.43.3748, PhysRevLett.76.388, *PhysRevLett.77.3995, *PhysRevLett.81.293, *PhysRevLett.83.1307, *PhysRevLett.91.123601, *Kiffner_review, *PhysRevLett.106.033001} and to examples of \mbox{x-ray} pulse shaping such as, e.g., studies of electromagnetically induced transparency for x~rays \cite{PhysRevLett.98.253001, *Glover_Nature}, for which an optical field is used to control \mbox{x-ray} absorption. {In our scheme, the imprinting of the optical frequency comb onto the \mbox{x-ray} spectrum takes advantage of a driving \mbox{x-ray} field influencing the precision with which the position of the peaks in the \mbox{x-ray} frequency comb is known. This comb is valuable as a relative ``ruler,'' e.g., to bridge an energy difference between an \mbox{x-ray} reference level and an unknown \mbox{x-ray} frequency at high energies for which, owing to the inefficiency of HHG at high harmonic orders, \mbox{x-ray} frequency-comb generation via HHG-based methods would encounter significant obstacles \cite{Kohler2012159}.}

The paper is structured as follows. In Sec.~\ref{Theoretical model} we present our theory in terms of an ensemble of four-level systems used to model the driven particles. We analyze the properties of the coherent and incoherent parts of the spectrum of resonance fluorescence related to many-particle effects and to periodic driving. The four-level scheme is applied in Sec.~\ref{Results and discussion} to He-like $\mathrm{Be}^{2+}$ ions to predict a frequency comb in the coherent part of the spectrum centered on the atomic transition at $\sim$\,$120\unit{eV}$. Section~\ref{Conclusion} concludes the paper. Atomic units are used throughout unless otherwise stated.

\section{Theoretical model}
\label{Theoretical model}
\subsection{Four-level model}

In this section, we present the four-level model used throughout the paper. The experimental setup is displayed in Fig.~\ref{fig:TheModel}(a). An \mbox{x-ray} field $\boldsymbol{\mathcal{E}}_{\mathrm{X}}(\boldsymbol{r},\,t)$, an optical continuous-wave (cw) auxiliary field $\boldsymbol{\mathcal{E}}_{\mathrm{L}}(\boldsymbol{r},\,t)$, and a periodic train of optical pulses $\boldsymbol{\mathcal{E}}_{\mathrm{C}}(\boldsymbol{r},\,t)$ of an optical-frequency-comb laser, irradiate an ensemble of ions. The fields copropagate in the $y$~direction; at time $t$ and position $\boldsymbol{r}$, for $q\in\{\mathrm{X,\,L,\,C}\}$, the incident fields are given by
\begin{equation}
\boldsymbol{\mathcal{E}}_{q}(\boldsymbol{r},\,t) = \mathcal{E}_{q,0}({ t'})\cos{[\omega_{q}t + \varphi_{q}({ t'}) + \varphi_{q,0} - \boldsymbol{k}_{q}\cdot\boldsymbol{r}]}\,\hat{\boldsymbol{e}}_q,
\label{eq:electricfields}
\end{equation}
with envelope $\mathcal{E}_{q,0}(t)$, carrier frequency $\omega_{q}$, phase $\varphi_{q}(t)$, carrier-envelope phase (CEP) $\varphi_{q,0}$, wavevector $\boldsymbol{k}_{q} = (\omega_{q}/c)\,\hat{\boldsymbol{e}}_{y}$, and linear polarization vector $\hat{\boldsymbol{e}}_q$. The intensity is \cite{diels2006ultrashort}
\begin{equation}
I_{q} = \frac{|\mathcal{E}_{q,0}|^2}{ 8\pi\alpha},
\label{eq:intensita}
\end{equation}
with the speed of light {in vacuum}  $c$ and the fine-structure constant $\alpha = 1/c$. Furthermore, {$t' = t-y/v_{q}$ is the retarded time due to the propagation of the pulses along the coordinate $y$, with $v_q$ being the group velocity of the $q$th field. We assume that the \mbox{x-ray} field $\boldsymbol{\mathcal{E}}_{\mathrm{X}}(\boldsymbol{r},\,t)$ and the optical cw auxiliary field $\boldsymbol{\mathcal{E}}_{\mathrm{L}}(\boldsymbol{r},\,t)$ are linearly polarized in the $z$ direction, $\hat{\boldsymbol{e}}_{\mathrm{X}} = \hat{\boldsymbol{e}}_{\mathrm{L}} = \hat{\boldsymbol{e}}_z $, while the train of pulses $\boldsymbol{\mathcal{E}}_{\mathrm{C}}(\boldsymbol{r},\,t)$ giving rise to the optical frequency comb is linearly polarized in the $x$ direction, $\hat{\boldsymbol{e}}_{\mathrm{C}} = \hat{\boldsymbol{e}}_x $, where $\hat{\boldsymbol{e}}_x$, $\hat{\boldsymbol{e}}_y$, and $\hat{\boldsymbol{e}}_z$ are unit vectors in the $x$, $y$, and $z$ directions. {Finally, for  $q\in\{\mathrm{X,\,L,\,C}\}$ we set the CEPs $\varphi_{q,0} = 0$ and we assume a dilute-gas setting, such that the phase velocity of all electric fields, to a very good approximation, equals $c$, resulting in good phase matching \cite{arXiv:1203.4127, PhysRevA.78.043409, JPSJ.30.518}.} 

\begin{figure}[t]
\centering%
\includegraphics[width=\columnwidth, keepaspectratio]{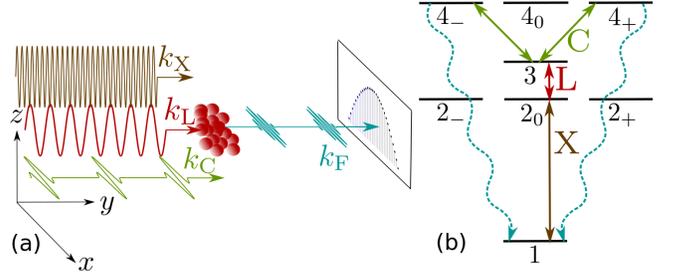}
\caption{(Color online) (a) An ensemble of ions is driven by {narrow-bandwidth} x~rays ($\boldsymbol{k}_{\mathrm{X}}$, brown), an auxiliary optical laser ($\boldsymbol{k}_{\mathrm{L}}$, red), both linearly polarized along the $z$~direction, and an optical frequency comb ($\boldsymbol{k}_{\mathrm{C}}$, green), linearly polarized along the $x$~direction. All fields propagate in the $y$~direction. The resonance fluorescence spectrum ($\boldsymbol{k}_{\mathrm{F}}$, blue) exhibits an induced \mbox{x-ray} frequency-comb structure. (b) Four-level scheme of $\mathrm{He}$-like ions interacting with the three light fields.}
\label{fig:TheModel}
\end{figure}

The bandwidth of $\boldsymbol{\mathcal{E}}_{\mathrm{L}}(\boldsymbol{r},\,t)$ is so small that it can be entirely neglected. We initially assume that $\boldsymbol{\mathcal{E}}_{\mathrm{X}}(\boldsymbol{r},\,t)$ has constant amplitude, $\mathcal{E}_{\mathrm{X},0}(t) = \bar{\mathcal{E}}_{\mathrm{X},0}$, and phase, $\varphi_{\mathrm{X}}(t) = 0$; the effect of the \mbox{x-ray} bandwidth is later taken into account by a stochastic approach \cite{PhysRevLett.37.1383}. Finally, the optical frequency comb has, to a very good approximation, constant phase $\varphi_{\mathrm{C}}(t) \equiv 0$ and a periodic amplitude 
\begin{equation}
\mathcal{E}_{\mathrm{{C}},0}(t) = \sum_{k=-\infty}^{+\infty} A_k\,\eu^{- \uimm \frac{2\pi k}{T_{\mathrm{p}}}t},
\end{equation}
with {repetition period $T_{\mathrm{p}}$ and} Fourier coefficients 
\begin{equation}
A_k = \frac{1}{T_{\mathrm{p}}}\,\int_{0}^{T_{\mathrm{p}}} \mathcal{E}_{\mathrm{{C}},0}(t) \, \eu^{\uimm \frac{2\pi k}{T_{\mathrm{p}}}t}\,\diff t.
\end{equation}
In other words, the envelope can be written as the following sum of identical pulses, 
\begin{subeqnarray}
\label{eq:opticalfrequencycomb}
\mathcal{E}_{\mathrm{{C},0}}(t) = \mathcal{E}_{\mathrm{{C},max}}\,\sum_{n = -\infty}^{+\infty} \mathcal{G}(t - nT_{\mathrm{p}}) , \\
\mathcal{G}(t) = \cos^2\biggl[ \frac{\pi}{T_{\mathrm{d}}} \Bigl(t - \frac{T_{\mathrm{d}}}{2}\Bigr)\biggr]\, \mathrm{R}\biggl[ \frac{1}{T_{\mathrm{d}}} \Bigl(t - \frac{T_{\mathrm{d}}}{2}\Bigr)\biggr] ,
\end{subeqnarray}
where, from Eq.~(\ref{eq:intensita}),
\begin{equation}
\mathcal{E}_{\mathrm{{C},max}} = \sqrt{8\pi\alpha I_{\mathrm{{C},max}}}
\end{equation} 
is the maximum electric-field strength of the train of optical pulses, associated with the maximum intensity $I_{\mathrm{{C},max}}$, and the rectangular function $\mathrm{R}(x)$ is defined in terms of the Heaviside step function $\theta(x)$ as 
\begin{equation}
\mathrm{R}(x) = \theta(x+1/2) - \theta(x-1/2). 
\end{equation}
The full width at half maximum (FWHM) of $\mathcal{G}^2(t)$ is \cite{0953-4075-42-23-235101}
\begin{equation}
T_{\mathrm{FWHM}} = 2\,T_{\mathrm{d}}\arccos{(\sqrt[4]{1/2})}/\pi,
\end{equation}
with $T_{\mathrm{d}}$ being the interval in which $\mathcal{G}(t)$ is different from 0, $T_{\mathrm{d}}\ll T_{\mathrm{p}}$.

The electric fields $\boldsymbol{\mathcal{E}}_{\mathrm{X}}(\boldsymbol{r},\,t)$, $\boldsymbol{\mathcal{E}}_{\mathrm{L}}(\boldsymbol{r},\,t)$, and $\boldsymbol{\mathcal{E}}_{\mathrm{C}}(\boldsymbol{r},\,t)$, drive isolated, electric-dipole (E1) transitions in the four-level system of Fig.~\ref{fig:TheModel}(b), where level $i$ has energy $\omega_i$ and the energy between levels $i$ and $j$ is given by 
\begin{equation}
\omega_{ij} = \omega_i - \omega_j,
\end{equation}
with $i,\,j\in S = \{1,\,2_{0,\pm},\,3,\,4_{0,\pm}\}$. The four-level model is applied to $\mathrm{He}$-like ions, with transition energies in the optical and \mbox{x-ray} ranges \cite{PhysRevA.81.022507}. In this case, $|1\rangle$ represents the ground state $1s^2$~$^1S_0$, with total-angular-momentum quantum number $J=0$ and positive parity. The states $|2_-\rangle$, $|2_0\rangle$, and $|2_+\rangle$, constitute the level $1s 2p$~$^3P_1$, with $J=1$ and negative parity. The quantum number $M_J$ associated with the $z$ component of the total-angular-momentum operator is, in the previous states, respectively equal to $-1$, $0$, and $1$. Furthermore, the state $|3\rangle$ is associated with the positive-parity level $1s 2s$~$^1S_0$, with $J=0$, whereas the three states $|4_-\rangle$, $|4_0\rangle$, and $|4_+\rangle$, represent the level $1s 2p$~$^1P_1$, with $J=1$ and negative parity \footnote{In $\mathrm{He}$-like ions, level $3$ has higher (lower) energy than level $2$ for a nuclear charge $Z\geq 7$ ($Z< 7$) \cite{PhysRevA.81.022507}.}. Other levels, such as $1s 2s$~$^3S_1$, $1s 2p$~$^3P_0$, and $1s 2p$~$^3P_2$, are not included in our description because they do not couple via an E1 interaction to the levels in Fig.~\ref{fig:TheModel}(b) and spontaneous-decay times from higher-energy levels to them are by orders of magnitude larger than the repetition period $T_{\mathrm{p}}$ of the optical frequency comb. 

All three excited levels, i.e., $2$, $3$, and $4$, are nonautoionizing, since, in all configurations, the passive electron remains in the $1s$ orbital, implying that the levels are energetically below the autoionizing threshold \cite{PhysRevA.81.022507}. Furthermore, other norm-nonconserving processes such as single-photon ionization due to $\boldsymbol{\mathcal{E}}_{\mathrm{X}}(\boldsymbol{r},\,t)$ \cite{Als-Nielsen:EM-01, *LANL} and multi-photon ionization due to $\boldsymbol{\mathcal{E}}_{\mathrm{C}}(\boldsymbol{r},\,t)$ \cite{1063-7869-47-9-R01, *Perelomov1, *Perelomov3} are safely negligible for the moderate, near-resonant fields employed here.

\subsection{Hamiltonian and equations of motion}
\label{Electric-dipole Hamiltonians}

The interaction of the electric fields $\boldsymbol{\mathcal{E}}_{q}(\boldsymbol{r},\,t)$ in Eq.~(\ref{eq:electricfields}), with $q\in\{\mathrm{X,\,L,\,C}\}$, and $N$ ions, respectively at positions $\boldsymbol{r}_n$, with $n\in\{1,\,\ldots,\,N\} $, is described by the Hamiltonian
\begin{equation}
\hat{H} = \hat{H}_{0} + \sum_{q \in\{\mathrm{X,\,L,\,C}\}}\hat{H}_{\mathrm{E1},q},
\label{eq:totalHamiltonian}
\end{equation}
where
\begin{equation}
\hat{H}_0 = \sum_{n=1}^N \sum_{i\in S}\omega_{i}\,\hat{\sigma}_{ii}^n
\end{equation}
is the atomic electronic structure Hamiltonian and 
\begin{equation}
\hat{H}_{\mathrm{E1},q} = \sum_{n=1}^N \hat{H}^n_{\mathrm{E1},q} = \sum_{n=1}^N\hat{\boldsymbol{d}}_n\cdot \boldsymbol{\mathcal{E}}_q(\boldsymbol{r}_n,\,t)
\label{eq:E1Hamiltonian}
\end{equation}
are the E1 interaction Hamiltonians \cite{Scully:QuantumOptics, johnson2007atomic}. In the previous equations, 
\begin{equation}
\hat{\sigma}_{ij}^n = |i\rangle_n\,_n\langle j|
\end{equation}
are the ladder operators, where $i,\,j \in S$ and $n\in\{1,\ldots,\,N\}$, while $\hat{\boldsymbol{d}}_n$ represents the dipole operator of an ion at position $\boldsymbol{r}_n$, 
\begin{equation}
\hat{\boldsymbol{d}}_n  = \sum_{i,j\in S} \boldsymbol{d}_{ij,n} \, \hat{\sigma}^n_{ij} ,
\label{eq:atomicoperator}
\end{equation}
with matrix elements
\begin{equation}
\boldsymbol{d}_{ij,n} =\,_n\langle i| \hat{\boldsymbol{d}}_n |j\rangle_n.
\end{equation}
Since the dipole moment is a property of the ion species, i.e., of the atomic number and the charge of the ion, the matrix elements $\boldsymbol{d}_{ij,n}= \boldsymbol{d}_{ij}$ do not explicitly depend on $n$. Furthermore, because $\hat{\boldsymbol{d}}_n$ is an irreducible tensor operator of rank 1 \cite{johnson2007atomic}, its vector components $\boldsymbol{d}_{ij}$, with $i,\,j \in S$, can be written as
\begin{equation}
{\boldsymbol{d}}_{ij} = {d}^{\,-1}_{ij}\hat{\boldsymbol{e}}_{\sigma_-} + {d}^{\,0}_{ij}\hat{\boldsymbol{e}}_{z} + {d}^{\,1}_{ij}\hat{\boldsymbol{e}}_{\sigma_+},
\label{eq:rank1operator}
\end{equation}
where, in addition to the Cartesian unit vectors $\hat{\boldsymbol{e}}_{x}$, $\hat{\boldsymbol{e}}_{y}$, and $\hat{\boldsymbol{e}}_{z}$, we define the circular-polarization vectors
\begin{equation}
\hat{\boldsymbol{e}}_{\sigma_{\pm}} = {(\mp\hat{\boldsymbol{e}}_x + \uimm\hat{\boldsymbol{e}}_y)}/{\sqrt{2}},
\label{eq:polarizationvec}
\end{equation}
with the positive or negative sign for polarizations $\lambda = \pm 1$. These are complex vectors, $\hat{\boldsymbol{e}}_{\sigma_{\pm}}^* = -  \hat{\boldsymbol{e}}_{\sigma_{\mp}}$, satisfying the orthogonality relations $\hat{\boldsymbol{e}}_{\sigma_{\pm}}\cdot \hat{\boldsymbol{e}}_{\sigma_{\pm}}^* = 1$ and $\hat{\boldsymbol{e}}_{\sigma_{\pm}}\cdot \hat{\boldsymbol{e}}_{\sigma_{\mp}}^* = 0$.
From parity considerations, the component $d^{\,k}_{ij}$ in Eq.~(\ref{eq:rank1operator}), with $i,\,j \in S$ and $k\in\{0,\pm1\}$, does not vanish only if $k$ is equal to the difference $M_{J,i} - M_{J,j}$ between the angular-momentum quantum numbers of the states $i$ and $j$ \cite{johnson2007atomic}. 

The three driving fields are assumed to be tuned to the respective transition energies, i.e., $\omega_{\mathrm{X}} = \omega_{21}$, $\omega_{\mathrm{L}} = |\omega_{32}|$, and $\omega_{\mathrm{C}} = \omega_{43}$. The effect of a field on the transitions to which it is not tuned is negligible \footnote{For example, the effect of the \mbox{x-ray} driving on the E1-allowed $1 \leftrightarrow 4_0$ transition can be, in the case of $\mathrm{Be}^{2+}$ ions, safely neglected, as it corresponds to a detuning of $\varDelta = 1.8\,\unit{eV}$, whereas the natural decay width of the excited state is $\varGamma_{41} = 5.05\times10^{-4}\unit{eV}$ and the assumed \mbox{x-ray} bandwidth $\gamma_{\mathrm{c}}$ is smaller than $2\pi/T_{\mathrm{p}} =4.14\times10^{-6}\,\unit{eV}$.} and the relevant interactions are highlighted in Fig.~\ref{fig:TheModel}(b). The states $|2_{\pm}\rangle$ and $|4_{0}\rangle$ are neglected, because they are not driven and the decay from higher-energy levels to them is orders of magnitude smaller than to the ground state.

With the previously described assumptions and in the rotating-wave approximation \cite{Scully:QuantumOptics}, the Hamiltonian $\hat{H}_{\mathrm{E1,X}}^n$ [Eq.~(\ref{eq:E1Hamiltonian})] describing the interaction of the $n$th ion with the \mbox{x-ray} field $\boldsymbol{\mathcal{E}}_{\mathrm{X}}(\boldsymbol{r},\,t) $ in Eq.~(\ref{eq:electricfields}) 
is given by
\begin{equation}
\hat{H}_{\mathrm{E1,X}}^n = \frac{d_{12_0}\bar{\mathcal{E}}_{\mathrm{X},0}}{2} \hat{\sigma}^n_{12_0}\,\eu^{\uimm(\omega_{\mathrm{X}}t - \boldsymbol{k}_{\mathrm{X}}\cdot\boldsymbol{r}_n)} \,+\,\mathrm{H.c.}
\label{eq:Hamiltonian12}
\end{equation}
The cw optical field $\boldsymbol{\mathcal{E}}_{\mathrm{L}}(\boldsymbol{r},\,t)$ 
interacts with the $n$th ion in a completely similar way \footnote{In the following discussion, we assume that $\omega_{32}>0$, this is, however, not an essential element of the derivation and the following calculations can be easily modified for $\omega_{32}<0$ (as necessary in order to apply our theory to $\mathrm{Be}^{2+}$ ions).}.
Finally, the train of optical pulses
\begin{equation}
\boldsymbol{\mathcal{E}}_{\mathrm{C}}(\boldsymbol{r},\,t)  = \mathcal{E}_{\mathrm{C},0}(t -  \boldsymbol{r}\cdot\hat{\boldsymbol{e}}_y/v_{\mathrm{C}})\,\cos{(\omega_{\mathrm{C}}t - \boldsymbol{k}_{\mathrm{C}}\cdot\boldsymbol{r})}\,\hat{\boldsymbol{e}}_x,
\end{equation}
tuned to the transition $|3\rangle \rightarrow |4_{\pm}\rangle$ and linearly polarized along the $\hat{\boldsymbol{e}}_x$ direction, $\hat{\boldsymbol{e}}_x = (\hat{\boldsymbol{e}}^*_{\sigma_{-}} - \hat{\boldsymbol{e}}^*_{\sigma_+})/\sqrt{2}$, interacts with the $n$th ion via the E1 interaction Hamiltonian 
\begin{equation}
\begin{aligned}
\hat{H}_{\mathrm{E1,C}}^n  =\,&\frac{1}{2} \sum_{j\in\{\pm\}}\hat{\sigma}^n_{34_{j}}\, \boldsymbol{d}_{34_{j}}\,\cdot\,\frac{\hat{\boldsymbol{e}}^*_{\sigma_{-}} - \hat{\boldsymbol{e}}^*_{\sigma_+}}{\sqrt{2}}  \\
\,&\times\,  \mathcal{E}_{\mathrm{C},0}(t -  \boldsymbol{r}_n\cdot\hat{\boldsymbol{e}}_y/{v_{\mathrm{C}}})\,\eu^{\uimm(\omega_{\mathrm{C}}t -\boldsymbol{k}_{\mathrm{C}}\cdot\boldsymbol{r}_n)}\,+\,\mathrm{H.c.},
\end{aligned}
\label{eq:HC1}
\end{equation}
with $\boldsymbol{d}_{34_{\pm}} = d^{\mp 1}_{34_{\pm}}\,\hat{\boldsymbol{e}}_{\sigma_{\mp}}$. Here, $d^{- 1}_{34_{+}}$ and $d^{+ 1}_{34_{-}}$ are the matrix elements of the electric-dipole momentum operator $\hat{\boldsymbol{d}}_n$ [Eq.~(\ref{eq:rank1operator})], which are related via the Clebsch-Gordan coefficients \cite{johnson2007atomic}. In this case, the explicit calculation of the Clebsch-Gordan coefficients allows one to observe that $d^{- 1}_{34_{+}} / d^{+ 1}_{34_{-}} =1$ and, therefore, to define the constant $\tilde{d}_{34} = d^{- 1}_{34_{+}} = d^{+ 1}_{34_{-}}$, in terms of which we rewrite the E1 interaction Hamiltonian~(\ref{eq:HC1}) as
\begin{equation}
\begin{aligned}
\hat{H}_{\mathrm{E1,C}}^n  =\,& \frac{\tilde{d}_{34}\mathcal{E}_{\mathrm{C},0}(t-  \boldsymbol{r}_n\cdot\hat{\boldsymbol{e}}_y/{v_{\mathrm{C}}})}{2}\\
\,&\times\,\sum_{j\in\{-1,\,+1\}} \frac{j}{\sqrt{2}}\hat{\sigma}^n_{34_j}\,\eu^{\uimm(\omega_{\mathrm{C}}t -\boldsymbol{k}_{\mathrm{C}}\cdot\boldsymbol{r}_n)}\,+\,\mathrm{H.c.}
\end{aligned}
\label{eq:Hamiltonian34}
\end{equation}


The ensemble of ions driven by the external fields is described via the density operator $\hat{\rho}(t)$, with elements $\rho_{ij}^n(t) = \langle\hat{\sigma}^n_{ji}(t)\rangle$, where $\langle \cdots \rangle$ stands for the expectation value of a quantum operator. The time evolution of the density operator is given by the Liouville--von Neumann equation  with system-reservoir interaction, i.e., by the master equation \cite{PhysRevLett.76.388, *PhysRevLett.77.3995, *PhysRevLett.81.293, *PhysRevLett.83.1307, *PhysRevLett.91.123601, *Kiffner_review, *PhysRevLett.106.033001, Scully:QuantumOptics}
\begin{equation}
\frac{\diff \hat{\rho}}{\diff t} = - \uimm\, [\hat{H}, \hat{\rho}] + \mathcal{L}[\hat{\rho}], 
\label{eq:master}
\end{equation}
where $\hat{H}$ is the Hamiltonian~(\ref{eq:totalHamiltonian}) and the Lindblad operator $\mathcal{L}[\hat{\rho}]$ represents the norm-conserving spontaneous decay of the system,
\begin{equation}
\mathcal{L}[\hat{\rho}] = \sum_{\substack{i,\,j\in S \\ \omega_{i} < \omega_{j}}} \sum_{n=1}^{N} -\frac{\varGamma_{ji}}{2}(\hat{\sigma}^n_{ji}\hat{\sigma}^n_{ij} \hat{\rho} \, -\, \hat{\sigma}^n_{ij}\hat{\rho}\hat{\sigma}^n_{ji} ) +\,\mathrm{H.c.},
\end{equation}
where the decay rates are given by $\varGamma_{ji} = 4\omega_{ji}^3\alpha^3|{\boldsymbol{d}}_{ij}|^2/3$ \cite{Scully:QuantumOptics}, with $i,\,j\in S$. Norm-nonconserving terms such as those from autoionization {or photoionization} are not present in our situation involving moderate, near-resonant fields.
The equations of motion (EOMs) from Eq.~(\ref{eq:master}), satisfied by the matrix elements $\rho_{ij}^n(t) = \langle\hat{\sigma}^n_{ji}(t)\rangle$ of the density operator $\hat{\rho}(t)$, can be more easily solved by introducing the slowly varying operators \cite{Scully:QuantumOptics, PhysRevA.22.2098, PhysRevA.45.4706, *PhysRevA.52.525} 
\begin{subeqnarray}
\hat{\varsigma}^n_{2_01}(t) &= &\hat{\sigma}^n_{2_01}(t)\,\eu^{-\uimm(\omega_{21}t - \boldsymbol{k}_{\mathrm{X}}\cdot \boldsymbol{r}_n)},\\ 
\hat{\varsigma}^n_{32_0}(t) &= &\hat{\sigma}^n_{32_0}(t)\,\eu^{-\uimm(\omega_{32}t - \boldsymbol{k}_{\mathrm{L}}\cdot \boldsymbol{r}_n)},\\
\hat{\varsigma}^n_{31}(t) &= &\hat{\sigma}^n_{31}(t)\,\eu^{-\uimm[\omega_{31}t - (\boldsymbol{k}_{\mathrm{X}} + \boldsymbol{k}_{\mathrm{L}})\cdot \boldsymbol{r}_n]},\\
\hat{\varsigma}^n_{4_{\pm}3}(t) &=& \hat{\sigma}^n_{4_{\pm}3}(t)\,\eu^{-\uimm(\omega_{43}t - \boldsymbol{k}_{\mathrm{C}}\cdot \boldsymbol{r}_n)},\\
\hat{\varsigma}^n_{4_{\pm}2_0}(t) &= &\hat{\sigma}^n_{4_{\pm}2_0}(t)\,\eu^{-\uimm[\omega_{42}t - (\boldsymbol{k}_{\mathrm{L}} + \boldsymbol{k}_{\mathrm{C}})\cdot \boldsymbol{r}_n]},\\
\hat{\varsigma}^n_{4_{\pm}1}(t) &= &\hat{\sigma}^n_{4_{\pm}1}(t)\,\eu^{-\uimm[\omega_{41}t - (\boldsymbol{k}_{\mathrm{X}} + \boldsymbol{k}_{\mathrm{L}} + \boldsymbol{k}_{\mathrm{C}}) \cdot \boldsymbol{r}_n]},
\label{eq:slowlyvarying}
\end{subeqnarray}
and $\hat{\varsigma}^n_{ij}(t) = [\hat{\varsigma}^n_{ji}(t)]\daga$. With these operators, we introduce the slowly varying density operator $\hat{\varrho}(t)$ of elements 
\begin{equation}
\varrho_{ij}^n(t) = \langle\hat{\varsigma}^n_{ji}(t)\rangle.
\end{equation}
From Eqs.~(\ref{eq:master}) and (\ref{eq:slowlyvarying}), the EOMs satisfied by the matrix elements $\varrho_{ij}^n(t)$ of the slowly varying density operator $\hat{\varrho}(t)$ are a set of coupled, linear differential equations with time-dependent coefficients. However, the only coefficients in the EOMs which explicitly depend on time are those associated with the envelope $\mathcal{E}_{\mathrm{C},0}(t-  \boldsymbol{r}_n\cdot\hat{\boldsymbol{e}}_y/{v_{\mathrm{C}}})$ of the pulse train $\boldsymbol{\mathcal{E}}_{\mathrm{C}}(\boldsymbol{r},\,t)$ appearing in Eq.~(\ref{eq:Hamiltonian34}). From Eq.~(\ref{eq:master}) it follows that the matrix elements of two density operators $\hat{\varrho}^n(t)$ and $\hat{\varrho}^{n'}(t)$, respectively associated with two ions at positions $\boldsymbol{r}_n$ and $\boldsymbol{r}_{n'}$, assume the same values at different times, i.e.,
\begin{equation}
\varrho_{ij}^n(t) = \varrho_{ij}^{n'}[t + (\boldsymbol{r}_{n'} -\boldsymbol{r}_n )\cdot\hat{\boldsymbol{e}}_y/v_{\mathrm{C}}],
\label{eq:ritardo}
\end{equation}
where the retardation effect displayed in Eq.~(\ref{eq:ritardo}) is due to the propagation of the train of pulses $\boldsymbol{\mathcal{E}}_{\mathrm{C}}(\boldsymbol{r},\,t)$ through the ensemble of particles.

Finally, we note that, for the cw~driving fields $\boldsymbol{\mathcal{E}}_{\mathrm{X}}(\boldsymbol{r},\,t)$ and $\boldsymbol{\mathcal{E}}_{\mathrm{L}}(\boldsymbol{r},\,t)$ and the periodic train of optical pulses $\boldsymbol{\mathcal{E}}_{\mathrm{C}}(\boldsymbol{r},\,t)$, the set of coupled, linear differential equations from~(\ref{eq:master}) exhibits periodic, time-dependent coefficients, with the repetition period $T_{\mathrm{p}}$ of the pulse train $\boldsymbol{\mathcal{E}}_{\mathrm{C}}(\boldsymbol{r},\,t)$. As a result, the master equation~(\ref{eq:master}) admits a periodic solution, in the following denoted as $\hat{\varrho}^{\mathrm{eq}}(t) = \hat{\varrho}^{\mathrm{eq}}(t+ T_{\mathrm{p}})$, which is asymptotically reached after turn-on effects have ceased.

%
%
%
%

\subsection{Spectrum of resonance fluorescence}
\label{Many-atom spectrum of resonance fluorescence from periodic EOMs}

While not coherently driven, the E1-allowed transition $4\rightarrow 1$ undergoes spontaneous decay [Fig.~\ref{fig:TheModel}(b)]: these photons, decaying from the states $|4_{\pm}\rangle$ with $M_J\in\{+1,\,-1\}$ to the state $|1\rangle$ with $M_J=0$, differ in energy and polarization from those decaying via the $2\rightarrow 1$ transition, i.e., they occupy a different region of the spectrum and can be distinguished via a polarization-dependent detector. In the following, therefore, we are allowed to directly focus on the elements of the emitted electric field and of the ensuing spectrum of resonance fluorescence which are associated with photon emission from the $4\rightarrow 1$ transition. Furthermore, we show that, by writing an optical frequency comb onto the driving cw \mbox{x-ray} field, a scheme connected to four-wave mixing \cite{Scully:QuantumOptics} is developed to imprint a frequency comb onto the coherent part of the spectrum in the forward direction. 

%
%

In order to calculate the spectrum of resonance fluorescence emitted by a set of driven ions, we introduce the electric-field operator
\begin{equation}
\hat{\boldsymbol{E}}(\boldsymbol{r},\,t) = \hat{\boldsymbol{E}}^+(\boldsymbol{r},\,t) + \hat{\boldsymbol{E}}^-(\boldsymbol{r},\,t),
\end{equation}
with positive- and negative-frequency components $\hat{\boldsymbol{E}}^+(\boldsymbol{r},\,t)$ and $\hat{\boldsymbol{E}}^-(\boldsymbol{r},\,t)$, respectively, where $\hat{\boldsymbol{E}}^-(\boldsymbol{r},\,t) = [\hat{\boldsymbol{E}}^+(\boldsymbol{r},\,t)]\daga$. The driven ions behave like oscillating dipole moments, emitting waves with a dipole radiation pattern. In the far-field region, the electric-field operator can be related to the dipole response of the atomic system via the following equality \cite{Scully:QuantumOptics}:
\begin{equation}
\begin{aligned}
\hat{\boldsymbol{E}}^+(\boldsymbol{r}, \, t) = \,&
\sum_{j\in \{4_+,\,4_-\}}\biggl[\ \frac{\omega_{41}^2}{c^2\,r}\,[{\boldsymbol{d}}_{j1} - \hat{\boldsymbol{e}}_r\,({\boldsymbol{d}}_{j1}\cdot \hat{\boldsymbol{e}}_r)]\\
\,&\times\,\sum_{n=1}^N \hat{\sigma}^n_{1j}(t - |\boldsymbol{r}- \boldsymbol{r}_n|/c)\biggr].
\label{eq:totalelectricfield}
\end{aligned}
\end{equation}
Here, $\boldsymbol{r} = r\,\hat{\boldsymbol{e}}_r$ is the detection point with respect to the ensemble of ions, at distance $r$ and along the observation direction given by the unit vector $\hat{\boldsymbol{e}}_r$.

The power spectrum of resonance fluorescence \cite{Scully:QuantumOptics} is given by
\begin{equation}
\begin{aligned}
S(\boldsymbol{r}, \omega) =&\, \frac{1}{4\pi^2\alpha}\lim_{T\rightarrow\infty}\frac{1}{T}\int_{-T/2}^{T/2}\int_{-T/2}^{T/2} \\
\,&\times\bigl[\bigl\langle \hat{\boldsymbol{E}}^-(\boldsymbol{r},\,t_1)\cdot \hat{\boldsymbol{E}}^+(\boldsymbol{r},\,t_2) \bigr\rangle\eu^{-\uimm\omega (t_2 - t_1)}\bigr]\diff t_1\diff t_2,
\end{aligned}
\label{eq:spectrum}
\end{equation}
where $S(\boldsymbol{r}, \omega)\,\diff \omega\,r^2\diff\varOmega$ represents the power emitted into the energy interval $[\omega,\,\omega+\diff\omega]$ and detected in a solid angle $\diff\varOmega$ centered at the observation point $\boldsymbol{r}$. Because of Eqs.~(\ref{eq:totalelectricfield}) and (\ref{eq:spectrum}), the spectrum depends on the two-time atomic expectation values $\langle\hat{\sigma}^n_{j1}(t_1)\,\hat{\sigma}^{n'}_{1j'}(t_2)\rangle$ \cite{Scully:QuantumOptics}, with $j,\,j'\in\{4_+,\,4_-\}$. Two contributions can thus be distinguished: a coherent spectrum $S^{\mathrm{coh}}(\boldsymbol{r}, \omega)$, depending on the product of single-time expectation values $\langle\hat{\sigma}^n_{j1}(t_1)\rangle\,\langle\hat{\sigma}^{n'}_{1j'}(t_2)\rangle$, and an incoherent spectrum $S^{\mathrm{inc}}(\boldsymbol{r}, \omega)$, related to $\langle \updelta\hat{\sigma}^n_{j1}(t_1)\,\updelta\hat{\sigma}^{n'}_{1j'}(t_2)\rangle$, where $\updelta\hat{\sigma}_{1j}^n = \hat{\sigma}_{1j}^n - \langle\hat{\sigma}_{1j}^n\rangle$.

In the following, we calculate the coherent part of the spectrum of resonance fluorescence which is emitted by the ensemble of ions in the forward direction, with $\hat{\boldsymbol{e}}_r\approx \hat{\boldsymbol{e}}_y$. 
Because of the emitted electric-field operator given in Eq.~(\ref{eq:totalelectricfield}), the spectrum of resonance fluorescence~(\ref{eq:spectrum}) exhibits position-dependent prefactors given by
\begin{equation}
[{\boldsymbol{d}}_{j1} - \hat{\boldsymbol{e}}_r\,({\boldsymbol{d}}_{j1}\cdot \hat{\boldsymbol{e}}_r)]\ [{\boldsymbol{d}}_{1j'} - \hat{\boldsymbol{e}}_r\,({\boldsymbol{d}}_{1j'}\cdot \hat{\boldsymbol{e}}_r)],
\label{eq:productss}
\end{equation}
with $j,\,j'\in\{4_+,\,4_-\}$, where the dipole-moment matrix elements $\boldsymbol{d}_{j1}$ and $\boldsymbol{d}_{1j'}$ are rank-1 tensors given by Eq.~(\ref{eq:rank1operator}). In particular, a close inspection of the associated Clebsch-Gordan coefficients allows one to notice that
\begin{subeqnarray}
\boldsymbol{d}_{4_+1} = \tilde{d}_{41}\,\hat{\boldsymbol{e}}_{\sigma^+}, &\boldsymbol{d}_{4_-1} = \tilde{d}_{41}\,\hat{\boldsymbol{e}}_{\sigma^-},\\
\boldsymbol{d}_{14_+} = -\tilde{d}_{41}^*\,\hat{\boldsymbol{e}}_{\sigma^-}, &\boldsymbol{d}_{14_-} = -\tilde{d}_{41}^*\,\hat{\boldsymbol{e}}_{\sigma^+},
\label{eq:dtildes}
\end{subeqnarray}
where $\tilde{d}_{41}$ is the amplitude of the dipole-moment matrix element and the circular-polarization vectors $\hat{\boldsymbol{e}}_{\sigma^{\pm}}$ were defined in Eq.~(\ref{eq:polarizationvec}). For observation directions $\hat{\boldsymbol{e}}_r$ close to the forward direction $\hat{\boldsymbol{e}}_y$ along which the three incident fields propagate, Eqs.~(\ref{eq:totalelectricfield}) and (\ref{eq:dtildes}), together with the definition of the resonance fluorescence spectrum~(\ref{eq:spectrum}), lead to
\begin{widetext}
\begin{equation}
\begin{aligned}
S^{\mathrm{coh}}(\boldsymbol{r}, \omega) =&  \frac{\omega_{41}^4\,|\tilde{d}_{41}|^2}{8\pi^2 c^3 r^2}\,\sum_{j,\,j'\in\{4_+,\,4_-\}}  \sum_{n=1}^N \sum_{n'=1}^N \lim_{T\rightarrow\infty}\frac{1}{T}\int_{-T/2}^{T/2}\int_{-T/2}^{T/2} \bigl[\bigl\langle \hat{\varsigma}^n_{j1}(t_1 - |\boldsymbol{r} - \boldsymbol{r}_n|/c)\bigr\rangle\,\bigl\langle\hat{\varsigma}^{n'}_{1j'}(t_2-|\boldsymbol{r} - \boldsymbol{r}_{n'}|/c)\bigr\rangle\\
&\times\,(-1)^{\delta_{jj'}+1}\,\eu^{-\uimm(\omega - \omega_{41})(t_2 - t_1)}\, \eu^{\uimm[(\omega_{41}/c)\,\hat{\boldsymbol{e}}_r -(\boldsymbol{k}_{\mathrm{X}}+ \boldsymbol{k}_{\mathrm{L}}+ \boldsymbol{k}_{\mathrm{C}})]\cdot(\boldsymbol{r}_n -\boldsymbol{r}_{n'})}\bigr]\,\diff t_1\,\diff t_2,
\end{aligned}
\label{eq:cohspectrumfirst}
\end{equation}
where $\delta_{jj'}$ is the Kronecker $\delta$~symbol. The position-dependent exponential function in the second line of Eq.~(\ref{eq:cohspectrumfirst}) renders the coherent part of the spectrum $S^{\mathrm{coh}}(\boldsymbol{r}, \omega)$ nonvanishing only in a small solid angle centered on $\hat{\boldsymbol{e}}_y$, as we discuss in the following [see, e.g., Eq.~(\ref{eq:eta})]. In this region, the space-dependent factors~(\ref{eq:productss}) do not display appreciable modifications and are therefore approximated with the value they exhibit at $\hat{\boldsymbol{e}}_r = \hat{\boldsymbol{e}}_y$. These factors, calculated by employing the definition of the complex polarization vectors $\hat{\boldsymbol{e}}_{\sigma^{\pm}}$ from Eq.~(\ref{eq:polarizationvec}), are responsible for the term $(-1)^{\delta_{jj'}+1}$ in the second line of Eq.~(\ref{eq:cohspectrumfirst}).

In order to proceed with the calculation of the spectrum of resonance fluorescence, we notice that the two states $|4_+\rangle$ and $|4_-\rangle$ are driven with opposite sign by the optical frequency comb. This is apparent from the factor $j$ in the E1 interaction Hamiltonian~(\ref{eq:Hamiltonian34}). As a result, the solution of the EOMs~(\ref{eq:master}) can be shown to satisfy $\varrho_{14_+}(t) = - \varrho_{14_-}(t)$, which can be employed to simplify the previously calculated spectrum~(\ref{eq:cohspectrumfirst}) to
\begin{equation}
\begin{aligned}
S^{\mathrm{coh}}(\boldsymbol{r}, \omega) =& 4 \frac{\omega_{41}^4\,|\tilde{d}_{41}|^2}{8\pi^2 c^3 r^2}\,\sum_{n=1}^N \sum_{n'=1}^N \lim_{T\rightarrow\infty}\frac{1}{T}\int_{-T/2}^{T/2}\int_{-T/2}^{T/2} \bigl[\bigl\langle \hat{\varsigma}^n_{4_+1}(t_1 - |\boldsymbol{r} - \boldsymbol{r}_n|/c)\bigr\rangle\,\bigl\langle\hat{\varsigma}^{n'}_{14_{+}}(t_2-|\boldsymbol{r} - \boldsymbol{r}_{n'}|/c)\bigr\rangle\\
&\times\,\eu^{-\uimm(\omega - \omega_{41})(t_2 - t_1)}\, \eu^{\uimm[(\omega_{41}/c)\,\hat{\boldsymbol{e}}_r -(\boldsymbol{k}_{\mathrm{X}}+ \boldsymbol{k}_{\mathrm{L}}+ \boldsymbol{k}_{\mathrm{C}})]\cdot(\boldsymbol{r}_n -\boldsymbol{r}_{n'})}\bigr]\,\diff t_1\,\diff t_2.
\end{aligned}
\label{eq:cohspectrum}
\end{equation}
The constructive interference among the four ``paths'' in Eq.~(\ref{eq:cohspectrumfirst}) leads to a reinforcement of the total spectrum, given by the factor four in Eq.~(\ref{eq:cohspectrum}). Similar interference effects in resonance fluorescence were described in Ref.~\cite{PhysRevLett.96.100403, *PhysRevA.73.063814}.

Although each ion emits independently of the other ones, in the forward direction, i.e., in the $\hat{\boldsymbol{e}}_y$ direction along which the three driving fields propagate, phase matching of emission from different ions is achieved \cite{arXiv:1203.4127}. This allows one to assume that $k_q = \omega_q/c$, for $q\in\{\mathrm{X,\,L,\,C}\}$. 
As a result, for $\boldsymbol{r} = r\, \hat{\boldsymbol{e}}_y$, the argument of the second exponential function in the second line of Eq.~(\ref{eq:cohspectrum}) vanishes, since $(\omega_{41}/c)\,\hat{\boldsymbol{e}}_r -(\boldsymbol{k}_{\mathrm{X}}+ \boldsymbol{k}_{\mathrm{L}}+ \boldsymbol{k}_{\mathrm{C}}) = 0$, and the spectrum reduces to 
\begin{equation}
\begin{aligned}
S^{\mathrm{coh}}(r\, \hat{\boldsymbol{e}}_y, \omega) =& \frac{\omega_{41}^4\,|\tilde{d}_{41}|^2}{2\pi^2 c^3 r^2}\,\sum_{n=1}^N \sum_{n'=1}^N\,\lim_{T\rightarrow\infty}\frac{1}{T}  \int_{-T/2}^{T/2}\int_{-T/2}^{T/2} \bigl[\bigl\langle \hat{\varsigma}^n_{4_+1}(t_1 - |\boldsymbol{r} - \boldsymbol{r}_n|/c)\bigr\rangle\,\bigl\langle \hat{\varsigma}^{n'}_{14_{+}}(t_2-|\boldsymbol{r} - \boldsymbol{r}_{n'}|/c)\bigr\rangle\\
&\times\,\eu^{-\uimm(\omega - \omega_{41})(t_2 - t_1)}\bigr]\,\diff t_1\,\diff t_2.
\end{aligned}
\label{eq:spectrumintermediate}
\end{equation}
In Eq.~(\ref{eq:spectrumintermediate}), the product of two complex conjugate terms can be recognized, which leads to
\begin{equation}
\begin{aligned}
S^{\mathrm{coh}}(r\, \hat{\boldsymbol{e}}_y, \omega) & = \frac{\omega_{41}^4\,|\tilde{d}_{41}|^2}{2\pi^2 c^3 r^2}\,\lim_{T\rightarrow\infty}\frac{1}{T}\,\left|\sum_{n=1}^N \int_{-T/2}^{T/2} \bigl\langle \hat{\varsigma}^n_{4_+1}(t_1 - |\boldsymbol{r} - \boldsymbol{r}_n|/c)\bigr\rangle\,
\eu^{\uimm(\omega - \omega_{41}) t_1}\,\diff t_1\right|^2\\
& = \frac{\omega_{41}^4\,|\tilde{d}_{41}|^2}{2\pi^2 c^3 r^2}\,N^2\,\lim_{T\rightarrow\infty}\frac{1}{T}\,\left|\int_{-T/2}^{T/2} \bigl\langle \hat{\varsigma}_{4_+1}(t)\bigr\rangle\,
\eu^{\uimm(\omega - \omega_{41}) t}\,\diff t\right|^2.
\end{aligned}
\label{eq:spectrumfinalforw}
\end{equation}

The just-described many-atom effect is essential to guarantee the directionality of the emitted coherent radiation. In the following, we describe the properties of the coherent part of the spectrum of resonance fluorescence for observation directions $\hat{\boldsymbol{e}}_r$ around the forward direction $\hat{\boldsymbol{e}}_y$ along which the three driving fields propagate. As discussed in Ref.~\cite{PhysRevA.45.4706, *PhysRevA.52.525}, the intensity of the coherent spectrum of resonance fluorescence rapidly falls for $\hat{\boldsymbol{e}}_r \neq \hat{\boldsymbol{e}}_y $, i.e., in a region for which the position-dependent factors~(\ref{eq:productss}) do not vary significantly and can therefore be assumed constant. From Eq.~(\ref{eq:cohspectrum}), this allows one to identify the rapidly varying, position-dependent term
\begin{equation}
\begin{aligned}
\eta(\hat{\boldsymbol{e}}_r) &= \frac{S^{\mathrm{coh}}(r\hat{\boldsymbol{e}}_r,\omega)}{S^{\mathrm{coh}}(r\hat{\boldsymbol{e}}_y,\omega)} = \Bigl|\frac{1}{N}\sum_{n=1}^N \eu^{\uimm\left[\frac{\omega_{41}}{c}\,\hat{\boldsymbol{e}}_r -(\boldsymbol{k}_{\mathrm{X}}+ \boldsymbol{k}_{\mathrm{L}}+ \boldsymbol{k}_{\mathrm{C}})\right]\cdot\boldsymbol{r}_n }\Bigr|^2 
\approx \Bigl|\frac{1}{L} \int_{-L/2}^{L/2}\eu^{\uimm \left[\frac{\omega_{41}}{c}\,(\hat{\boldsymbol{e}}_r -\hat{\boldsymbol{e}}_y)\right]\cdot \,(y \,\hat{\boldsymbol{e}}_y) }\diff y \Bigr|^2 \\
&= \sinc^2{\left\{\frac{\omega_{41} L }{2c}[\cos(\phi) - 1]\right\}}.
\end{aligned}
\label{eq:eta}
\end{equation}
\end{widetext}
In Eq.~(\ref{eq:eta}), $\phi$ is the angle that the unit vector $\hat{\boldsymbol{e}}_r$, associated with the direction of observation, forms with the $y$ axis, i.e., $\cos\phi = \hat{\boldsymbol{e}}_r \cdot \hat{\boldsymbol{e}}_y$, and $\sinc{(x)} = \sin{(x)}/x$. Furthermore, while going from the first to the second line in Eq.~(\ref{eq:eta}), we have approximated the sum over the $N$ ions with an integral over the coordinate $y = \boldsymbol{r}_n\cdot\hat{\boldsymbol{e}}_y$, assuming a length $L$ of the ion sample and a constant linear density $N/L$ \cite{PhysRevA.45.4706, *PhysRevA.52.525}. For $\phi = 0$, $\eta(\hat{\boldsymbol{e}}_r)$ is clearly equal to 1. However, the function $\eta(\hat{\boldsymbol{e}}_r)$ determines an emission cone with opening angle $\phi^*$. This is here defined as the angle corresponding to the first zero of Eq.~(\ref{eq:eta}), i.e., satisfying the identity 
\begin{equation}
\frac{\omega_{41} L }{2c}[\cos(\phi^*) - 1] = \frac{\pi}{2}.
\end{equation}
The resulting opening angle of the emission cone
\begin{equation}
\phi^* \approx \sqrt{\frac{2c \pi}{\omega_{41} L}}
\end{equation}
and the distance $r$ at which the spectrum is observed allow one to define the area 
\begin{equation}
\Delta A  = r^2 \int_{\Delta\varOmega} \diff \varOmega = \pi [r \sin(\phi^*)]^2 \approx \frac{2\pi^2 c r^2}{\omega_{41} L}
\label{eq:area}
\end{equation}
in the solid angle $\Delta\varOmega$ about the forward direction $\hat{\boldsymbol{e}}_y$ in which the radiation is emitted.

In contrast to the just-described part of the spectrum of resonance fluorescence, for which the coherent emission in the forward direction gives rise to a multiplication factor of $N^2$ in Eq.~(\ref{eq:spectrumfinalforw}), the incoherent part of the spectrum $S^{\mathrm{inc}}(r\, \hat{\boldsymbol{e}}_r, \omega)$  is only proportional to $N$ and completely lacks space-directionality contributions from many-particle effects \cite{PhysRevA.45.4706, *PhysRevA.52.525}. No terms such as $\eta(\hat{\boldsymbol{e}}_r)$ are present in the incoherent spectrum and the only position-dependent contribution is given by the terms in Eq.~(\ref{eq:productss}) \cite{PhysRevA.45.4706, *PhysRevA.52.525}. In the forward direction, the incoherent part of the spectrum is smaller than the coherent spectrum by a factor $N$ and will hence be neglected in the following.

We conclude this section by focusing on the effects on the coherent part of the spectrum of resonance fluorescence due to the periodicity of the EOMs obtained from the master equation~(\ref{eq:master}). As we previously mentioned, since the linear differential equations determining the time evolution of the density operators $\hat{\varrho}^n(t)$ have coefficients which are periodic in time, there exists a periodic solution $\hat{\varrho}^{\mathrm{eq}}(t)$ of the EOMs. This solution has the same period $T_{\mathrm{p}}$ as the repetition time of the train of pulses $\boldsymbol{\mathcal{E}}_{\mathrm{C}}(\boldsymbol{r},\,t)$ from the optical frequency comb which drives the ensemble of ions. When turn-on effects have ceased, any solution $\hat{\varrho}^{n}(t)$ converges to $\hat{\varrho}^{\mathrm{eq}}(t)$, independent of the initial state of the system. As discussed in Ref.~\cite{PhysRevA.22.2098}, we can take advantage of this periodic solution in Eq.~(\ref{eq:spectrumfinalforw}) to show that the coherent part of the spectrum emitted on the $4\rightarrow 1$ transition {in the forward direction} consists of an \mbox{x-ray} frequency comb centered on the frequency $\omega_{41}$ with the same tooth spacing as the driving optical frequency comb,
\begin{subeqnarray}
\label{eq:xraycomb}
\slabel{eq:S4114tot}
S^{\mathrm{coh}}(r\,\hat{\boldsymbol{e}}_y, \omega) =  \,\sum_{m=-\infty}^{+\infty} \mathcal{S}_m\, \delta\Bigl( \omega - \omega_{41} - \frac{2\pi m}{T_{\mathrm{p}}}\Bigr),\\
\slabel{eq:Sm}
\mathcal{S}_m = \frac{\omega_{41}^4\,|\tilde{d}_{41}|^2}{\pi c^3 r^2}\,N^2 \ \Bigl|\frac{1}{T_{\mathrm{p}}} \int_{0}^{T_{\mathrm{p}}}\varrho^{\mathrm{eq}}_{4_{+}1}(t)\,\eu^{\uimm \frac{2\pi m} {T_{\mathrm{p}}} t }\,\diff t\Bigr|^2.
\end{subeqnarray}
{Here, $\delta(x)$ is the Dirac $\delta$~function, $\varrho_{14_+}^{\mathrm{eq}}(t)$ the relevant matrix element of the periodic, slowly varying density operator $\hat{\varrho}^{\mathrm{eq}}(t)$, and $\tilde{d}_{41}$ was defined in Eq.~(\ref{eq:dtildes}). 
Because of {many-ion effects}, the photons emitted in the forward direction are focused in a beam whose mean area is given by Eq.~(\ref{eq:area}) \cite{PhysRevA.45.4706, *PhysRevA.52.525}. By recalling that the spectrum of resonance fluorescence is defined as the emitted power per unit area $A$ and unit frequency $\omega$, it follows that 
\begin{equation}
\begin{split}
P_m &= \int_{\Delta\varOmega} \int_{\omega_m - \pi/T_{\mathrm{p}}}^{\omega_{m} + \pi/T_{\mathrm{p}}} S^{\mathrm{coh}}(r\,\hat{\boldsymbol{e}}_r, \omega) \,\diff\omega \,r^2\,\diff\varOmega = \mathcal{S}_m \, \Delta A \\
&= \frac{ 2\pi  \omega_{41}^3\,|\tilde{d}_{41}|^2}{  L  c^2 }\,N^2 \, \Bigl|\frac{1}{T_{\mathrm{p}}} \int_{0}^{T_{\mathrm{p}}}\varrho^{\mathrm{eq}}_{4_{+}1}(t)\,\eu^{\uimm \frac{2\pi m} {T_{\mathrm{p}}} t }\,\diff t\Bigr|^2
\end{split}
\label{eq:Pm}
\end{equation}
describes the power of the $m$th peak in the spectrum at frequency $\omega_m =\omega_{41} + {2\pi m}/{T_{\mathrm{p}}} $.

\section{Results and discussion}
\label{Results and discussion}
In this section, we apply the previously described theoretical model to predict a frequency comb at \mbox{x-ray} frequencies. In particular, we aim at generating a comb with the same number of peaks, i.e., overall width, as the driving optical frequency comb, and with emitted power comparable to the power of present-day XUV combs generated via HHG \cite{Nature.482.68}. In order to bridge an energy difference between two \mbox{x-ray} levels, a sufficiently wide comb is needed. Furthermore, powers on the same order of magnitude guarantee that our predicted comb could be similarly detected and used as XUV combs from presently explored methods. 

In the following, we describe how we proceed to maximize the number and the power $P_m$ of the peaks in the comb~(\ref{eq:xraycomb}). From  Eq.~(\ref{eq:Pm}), the power associated with the $m$th peak is proportional to the modulus squared of the $m$th Fourier coefficient of the periodic function $\varrho^{\mathrm{eq}}_{4_{+}1}(t)$. The properties of a Fourier-series expansion \cite{arfken2011mathematical} imply that the overall width of the spectrum is inversely proportional to the duration of $\varrho_{4_{+}1}^{\mathrm{eq}}(t)$. In order to produce an \mbox{x-ray} frequency comb with as many teeth as in the driving optical frequency comb, the matrix element $\varrho_{4_{+}1}^{\mathrm{eq}}(t)$ needs to consist of pulses closely following the envelope ${\mathcal{E}}_{\mathrm{{C}},0}(t)$ of the pulse train $\boldsymbol{\mathcal{E}}_{\mathrm{C}}(\boldsymbol{r},\,t)$ of the driving optical-frequency-comb laser. 

This can be better understood by introducing the pulse area of a single pulse in the train $Q =  \int_{0}^{T_{\mathrm{p}}} |\boldsymbol{d}_{34_+}|\,{\mathcal{E}}_{\mathrm{{C}},0}(t)\,\diff t $. When the envelope of the driving pulse satisfies the condition $Q = 2n\pi$, the atomic variables of the system perform an integer number of Rabi cycles \cite{Scully:QuantumOptics} after which population and coherence of the highest level are brought back to $0$ exactly at the end of the pulse \cite{PhysRev.40.502, *PhysRevA.23.2496, *0022-3700-17-15-005, PhysRevA.17.247, *Lewenstein:86, *PhysRevA.86.033402}. Conversely, for $Q\neq 2n\pi$, the atomic variables are led to nonvanishing values at the end of the pulse, such that spontaneous decay of the highest level follows. By choosing the peak intensity $I_{\mathrm{{C}, max}}$ of the pulse train $\boldsymbol{\mathcal{E}}_{\mathrm{C}}(\boldsymbol{r},\,t)$ to fulfill the condition $Q = 2n \pi$, we guarantee that, in the absence of an optical pulse, the population of the states $|4_{\pm}\rangle$ and the off-diagonal terms ${\varrho}_{4_{\pm}1}^{\mathrm{eq}}(t)$ vanish exactly. The emitted spectrum [Eq.~(\ref{eq:xraycomb})] consists of peaks whose power, from Eq.~(\ref{eq:Pm}), is proportional to the Fourier coefficient of a function which is different from 0 only in the presence of the optical pulses, i.e., in an interval of duration $T_{\mathrm{FWHM}}$ given by the FWHM duration of the pulses giving rise to the optical frequency comb. As a result of the properties of a Fourier-series expansion, the overall width of the spectrum is thus given by $\sim$\,$2\pi/T_{\mathrm{FWHM}}$. Conversely, if the intensity of the pulse train $\boldsymbol{\mathcal{E}}_{\mathrm{C}}(\boldsymbol{r},\,t)$ is not properly chosen, every pulse from the optical-frequency-comb laser gives rise to a subsequent long decay of the atomic variables, ${\varrho}_{4_{+}1}^{\mathrm{eq}}(t) \sim \eu^{-\varGamma_{41} t}$, which affects the amplitude of the peaks in Eq.~(\ref{eq:Pm}) and, therefore, results in a spectrum of smaller width, $\varGamma_{41} \ll 2\pi/T_{\mathrm{FWHM}}$, and smaller number of relevant teeth.

\begin{figure}[tb]
\centering%
\includegraphics[width=\linewidth, keepaspectratio]{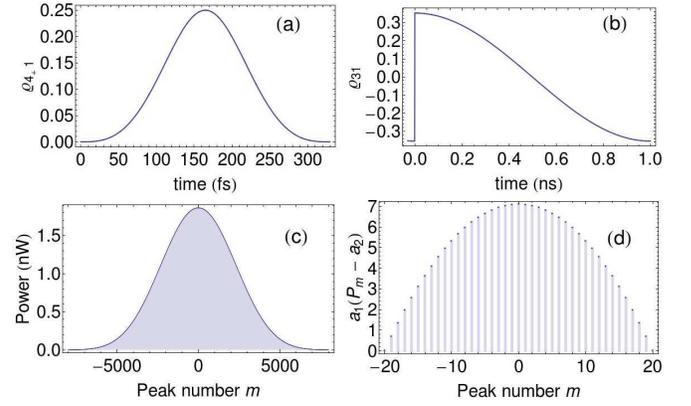}
\caption{(Color online) Time evolution of the {periodic} density operator $\hat{\varrho}^{\mathrm{eq}}(t)$ and spectrum of resonance fluorescence for $\mathrm{Be^{2+}}$ ions. Present-day parameters are used to model the optical frequency comb [Eq.~(\ref{eq:opticalfrequencycomb})], $T_{\mathrm{FWHM}} = 120\,\unit{fs}$, {$T_{\mathrm{p}} = 1\,\unit{ns}$, $1/T_{\mathrm{p}} = 1\,\unit{GHz}$} 
\cite{Nature.482.68, Hartl:07, *nphoton.2008.79, *Eidam:10, *Ruehl:10}, i.e., {$2\pi/T_{\mathrm{p}} = 4.1\times 10^{-6}\,\unit{eV}$}. 
The ion sample has $N = 10^6$ particles over a length of $L = 1\,\unit{cm}$ {and area $1\,\mathrm{mm^2}$}. The driving fields have intensities {$I_{\mathrm{X}} = 1.5\times10^{4}\,\unit{W/cm^2}$, $I_{\mathrm{L}} = 1.7\times10^8\,\unit{W/cm^2}$}, and $I_{\mathrm{C, max}} = 3.0\times10^{10}\,\unit{W/cm^2}$, associated with $2\pi$ optical-frequency-comb pulses. The periodic solutions are (a) $\varrho^{\mathrm{eq}}_{4_+1}(t)$ for $nT_{\mathrm{p}} < t < nT_{\mathrm{p}} + T_{\mathrm{d}}$, with $ \,T_{\mathrm{d}} = \pi T_{\mathrm{FWHM}} /[2\arccos{(\sqrt[4]{1/2})}]$, and (b) $\varrho^{\mathrm{eq}}_{31}(t)$ for $nT_{\mathrm{p}} < t < (n+1)T_{\mathrm{p}} $. The power $P_m$ of each peak in the spectrum of Eq.~(\ref{eq:xraycomb}) is displayed (c) for the whole comb, centered on $\omega_{41} = 123.7\,\unit{eV}$, and (d) around the maximum. In panel (d), $a_1 = 10^5\, \unit{nW^{-1}}$, {$a_2 = 1.86\,\unit{nW}$}.}%
\label{fig:HeLikeBeTotal}
\end{figure}

In Fig.~\ref{fig:HeLikeBeTotal} we show results obtained by applying our four-level scheme to model isolated transitions in $\mathrm{He}$-like $\mathrm{Be}^{2+}$ ions. The decay rates $\varGamma_{ji}$ are calculated with \texttt{grasp2K} \cite{DBLP:journals/cphysics/JonssonHFG07}, while the transition energies $\omega_{21} = 121.9\,\unit{eV}$, $\omega_{23} = 0.2699\,\unit{eV}$, $\omega_{43} = 2.018\,\unit{eV}$, and $\omega_{41} = 123.7\,\unit{eV}$, are taken from Ref.~\cite{PhysRevA.81.022507}. {We assume a density of $\mathrm{Be}^{2+}$ ions of $10^8\,\mathrm{cm^{-3}}$, which can be reached with an electron-beam ion trap \cite{PhysRevLett.107.143002, BernittNature}. For such a dilute sample, good phase matching is achieved \cite{arXiv:1203.4127, PhysRevA.78.043409, JPSJ.30.518}. Alternative experimental settings, e.g., by gas discharge or photoionization by an \mbox{x-ray} pre-pulse \cite{PhysRevLett.107.233001, Rohringer.Nature.481.2012}, may allow for higher densities, but one ought to ensure that a stable environment is obtained, such that all pulses in the optical frequency comb encounter a constant density of ions, atoms, and free electrons. This is discussed in the appendix}.

By using an optical frequency comb composed of $2\pi$ pulses, the matrix elements of the periodic density operator $\hat{\varrho}^{\mathrm{eq}}(t)$ related to the states $|4_{\pm}\rangle$ vanish after each pulse. The time evolution of $\varrho^{\mathrm{eq}}_{4_+1}(t)$ in the presence of an optical pulse is exhibited in Fig.~\ref{fig:HeLikeBeTotal}(a), where it is apparent that the vanishing initial value is reached again at the end of the interaction. In the interval in between two optical pulses, when the excited states $|4_{\pm}\rangle$ are completely depopulated, the remaining states $|1\rangle$, $|2_0\rangle$, and $|3\rangle$, behave like a three-level system \cite{PhysRevA.42.1630, *PhysRevA.43.3748} driven by the two cw fields $\boldsymbol{\mathcal{E}}_{\mathrm{X}}(\boldsymbol{r},\,t)$ and $\boldsymbol{\mathcal{E}}_{\mathrm{L}}(\boldsymbol{r},\,t)$. These fields stimulate oscillations of the remaining elements $\varrho^{\mathrm{eq}}_{ij}(t)$ of the density operator, with $i,\,j\in\{1,\,2_0,\,3\}$, as shown in Fig.~\ref{fig:HeLikeBeTotal}(b), and affect the periodic behavior of the entire density operator. In other words, the intensities $I_{\mathrm{X}}$ and $I_{\mathrm{L}}$ determine the amplitude of the oscillating function $\varrho^{\mathrm{eq}}_{31}(t)$ [Fig.~\ref{fig:HeLikeBeTotal}(b)], indirectly influencing also the peak value displayed by $\varrho_{4_+1}^{\mathrm{eq}}(t)$ in Fig.~\ref{fig:HeLikeBeTotal}(a). Given the relationship appearing in Eq.~(\ref{eq:Pm}) between the amplitude of $\varrho_{4_+1}^{\mathrm{eq}}(t)$ and the intensity of the peaks in the emitted \mbox{x-ray} frequency comb, it is important to properly set the peak intensities $I_{\mathrm{X}}$ and $I_{\mathrm{L}}$ in order to maximize the peak value of $\varrho_{4_+1}^{\mathrm{eq}}(t)$ and, consequently, the emitted photon number.

Having suppressed the post-pulse decay of $\varrho^{\mathrm{eq}}_{4_+1}(t)$ by choosing a train of $2\pi$-area pulses, the resulting spectrum of resonance fluorescence [Eq.(\ref{eq:xraycomb})] is shown in Fig.~\ref{fig:HeLikeBeTotal}(c); it is centered on $\omega_{41} = 123.7\,\unit{eV}$ and contains {$\sim$\,$10^4$} peaks with an energy spacing of {$2\pi/T_{\mathrm{p}} =4.1\times10^{-6}\,\unit{eV}$}. Figure~\ref{fig:HeLikeBeTotal}(d) highlights the comb structure of the spectrum. The peak intensity of $I_{\mathrm{{C}, max}} = 3.0\times10^{10}\,\unit{W/cm^2}$ is much lower than those needed for the generation of XUV frequency combs via HHG \cite{Hartl:07, *nphoton.2008.79, *Eidam:10, *Ruehl:10} and the power of each peak in the emitted spectrum \footnote{This also holds for lower repetition frequencies, $1/T_{\mathrm{p}}=100\,\mathrm{MHz}$, where peak powers of the order of tens of picowatts are predicted.} is comparable to the power which was measured in Ref.~\cite{Nature.482.68}.

The results presented so far were obtained by assuming cw x~rays with a vanishing bandwidth. To incorporate the effect of a finite bandwidth $\gamma_{\mathrm{c}}$ of the \mbox{x-ray} light source, we adopt the approach from Ref.~\cite{PhysRevLett.37.1383} which includes the influence of the temporal fluctuations of the driving field on the spectrum of resonance fluorescence. In this case, the \mbox{x-ray} field $\boldsymbol{\mathcal{E}}_{\mathrm{X}}(t)$ is a stochastic variable which varies in the ensemble of all possible realizations of the stochastic process. Thereby, it is possible to derive the EOMs for the ensemble-averaged density operator and thus obtain the ensemble-averaged spectrum of resonance fluorescence. By following this approach, one obtains an ensemble-averaged spectrum which is a continuous function still displaying peaks at frequencies $\omega_m = \omega_{41} + 2\pi m/T_{\mathrm{p}}$, as in Eq.~(\ref{eq:xraycomb}). However, because of the finite bandwidth $\gamma_{\mathrm{c}}$ of the driving \mbox{x-ray} field, the $\delta$ peaks exhibited by Eq.~(\ref{eq:xraycomb}) are broadened and each peak in the ensemble-averaged spectrum features a spectral FWHM of $\sim$\,$2\gamma_{\mathrm{c}}$. To preserve also for $\gamma_{\mathrm{c}}\neq 0$ the frequency-comb structure we predicted in Eq.~(\ref{eq:xraycomb}), we need to ensure that the spectral width of the teeth in the imprinted comb is lower than their separation energy. From the previous considerations, this implies that the \mbox{x-ray} bandwidth ought to be smaller than the repetition frequency of the optical frequency comb, i.e., {$2\gamma_{\mathrm{c}}< 2\pi/T_{\mathrm{p}} = 4.1\times10^{-6}\,\unit{eV}$}. The many-peak structure is otherwise washed out and the peaks in the spectrum cannot be clearly distinguished. X~rays with such a small bandwidth are not available at present. Yet, by increasing the repetition frequency $2\pi/T_{\mathrm{p}}$ of the optical frequency comb, a wider \mbox{x-ray}-comb tooth spacing would result and a larger \mbox{x-ray} bandwidth may be accommodated. {With a peak intensity of $I_{\mathrm{C,max}} = 3\times 10^{10}\,\mathrm{W/cm^2}$, such an increase in the repetition rate of the optical frequency comb is feasible \cite{PhysRevLett.94.193201, *Nature.436.234, Nature.482.68, Sander:12}.} Furthermore, we notice that the quality and the coherence of \mbox{x-ray} sources have dramatically improved during the last decades. Although present \mbox{x-ray} sources do not provide the resolving powers required here \cite{nphoton.2010.176, *nphoton.2007.76, *nphoton.2011.178, *RepXFEL}, new schemes \cite{nphoton.2012.180, Rohringer.Nature.481.2012, Gabor_paper, *Buth:11, PhysRevLett.100.244802} show the strong need for narrower-bandwidth x~rays and the remarkable attempts to reach them.

\section{Conclusions}
\label{Conclusion}
In this paper, we present an \mbox{x-ray} pulse-shaping method to directly access the time evolution of a driven atomic system and stimulate the periodic emission of x~rays via a three-color scheme in a four-level system. This is investigated by calculating the coherent part of the spectrum of resonance fluorescence which is emitted by an ensemble of ions in the forward direction. The model is applied to imprint an optical frequency comb onto cw x~rays. We employ $\mathrm{He}$-like $\mathrm{Be}^{2+}$ ions as an atomic implementation of the model. We show that a frequency comb is generated, which is centered on the \mbox{x-ray} transition energy at $123.7\,\unit{eV}$ and which requires peak intensities of the driving optical frequency comb which are lower by several orders of magnitude than those presently needed for HHG-based comb-generation methods \cite{Nature.482.68}. 

Although the four-level model developed in this paper was applied to He-like $\mathrm{Be}^{2+}$ ions, the scheme has general validity and can be employed to describe different systems with potentially higher \mbox{x-ray} transition energies. Similar results, for example, can be obtained from other $\mathrm{He}$-like ions, such as $\mathrm{Ne}^{8+}$. 
In this case, for $\omega_{41} = 922.0\,\unit{eV}$ \cite{PhysRevA.81.022507}, a comb in the keV~range can be predicted, 
yet for the transition energy $\omega_{43} = 6.679\,\unit{eV}$ intense optical frequency combs are not available. Our model can be also applied to different atomic transitions, e.g., $1s^2 \rightarrow 1s\,np$ with $n\geq 3$ in heavier ions, for which experimentally accessible \mbox{x-ray} and optical energies can be found; or even to nuclear transitions up to the $\gamma$~range. 

Our scheme takes advantage of narrow-bandwidth \mbox{x-ray} sources. We recognize that the assumption of a very narrow \mbox{x-ray} bandwidth does not allow an implementation of our scheme with currently available \mbox{x-ray} technology. Nevertheless, we are confident that the advances in \mbox{x-ray} science and the constant improvement in the quality and coherence of \mbox{x-ray} sources will soon provide the experimental conditions necessary to demonstrate the viability of the scheme. Not only does the model tackle the problem of \mbox{x-ray} comb generation, with the advantage of being applicable at energies for which existing methods would not be adequate, but it also represents an example of how the resonance fluorescence spectrum emitted by an ensemble of driven particles can be manipulated by directly controlling the time evolution of the atomic system.\\

\acknowledgments

S.M.C.~and Z.H.~acknowledge helpful discussions with J\"org~Evers, Christian~Ott, and Thomas~Pfeifer. The work of Z.H.~was supported by the Alliance Program of the Helmholtz Association (HA216/EMMI).

\appendix

\section{Phase-matching effects on the spectrum of resonance fluorescence}
\label{Phase-matching effects on the spectrum of resonance fluorescence}

Here, we discuss the influence of phase matching \cite{arXiv:1203.4127} on the shape of the spectrum of resonance fluorescence~(\ref{eq:cohspectrum}). Depending on the experimental setting used to produce the plasma of ions in our scheme, phase matching may have to be taken into account. In the derivation presented in Sec.~\ref{Theoretical model}, we have assumed that the presence of atoms, ions, and free electrons in the plasma does not influence the phase velocity $v_q = c/n_q$ of the three driving fields, where $n_q$ is the refractive index of the medium at the frequency of the $q$th field. At the very low atom number densities considered in Sec.~\ref{Results and discussion}, this is a valid approximation. However, one needs to quantify how phase-matching effects modify the predicted spectrum of resonance fluorescence, in case this represents an issue at higher densities than those assumed here. 

As done in Sec.~\ref{Many-atom spectrum of resonance fluorescence from periodic EOMs}, we consider here the coherent part of the spectrum of resonance fluorescence on the $|4_{\pm}\rangle\rightarrow |1\rangle$ transition in the forward direction. By introducing \cite{arXiv:1203.4127}
\begin{equation}
\begin{aligned}
\Delta k &\, = \frac{\omega_{41}}{c} -(|\boldsymbol{k}_{\mathrm{X}}|+ |\boldsymbol{k}_{\mathrm{L}}|+ |\boldsymbol{k}_{\mathrm{C}}|)\\
&\, =\frac{1}{c} (\omega_{41} - \omega_{21} n_{\mathrm{X}} - \omega_{32} n_{\mathrm{L}} - \omega_{43} n_{\mathrm{C}}),
\end{aligned}
\end{equation}
the following spectrum can be obtained from Eq.~(\ref{eq:cohspectrum}),
\begin{equation}
\begin{aligned}
& S^{\mathrm{coh}}(r\hat{\boldsymbol{e}}_y, \omega) \\
&\ \,= \frac{\omega_{41}^4|\tilde{d}_{41}|^2}{2\pi^2 c^3 r^2}\lim_{T\rightarrow\infty}\frac{1}{T}\biggl|\sum_{n=1}^N \int_{-T/2}^{T/2} \bigl\langle \hat{\varsigma}^n_{4_+1}(t_1 - |\boldsymbol{r} - \boldsymbol{r}_n|/c)\bigr\rangle\\
&\ \ \ \ \ \times\,\eu^{\uimm(\omega - \omega_{41}) t_1}\,\eu^{\uimm\,\Delta k\,\hat{\boldsymbol{e}}_y\cdot\boldsymbol{r}_n }\,\diff t_1\biggr|^2,
\end{aligned}
\label{eq:spectrumforapp}
\end{equation}
where the same steps were followed which we described in Sec.~\ref{Many-atom spectrum of resonance fluorescence from periodic EOMs} while proceeding from Eq.~(\ref{eq:cohspectrum}) to Eq.~(\ref{eq:spectrumfinalforw}). We note in particular that, for $\Delta k = 0$, Eq.~(\ref{eq:spectrumforapp}) leads exactly to Eq.~(\ref{eq:spectrumfinalforw}).

The additional exponential function $\eu^{\uimm\,\Delta k\,\hat{\boldsymbol{e}}_y\cdot\boldsymbol{r}_n }$ appearing in Eq.~(\ref{eq:spectrumforapp}) implies that, in the forward direction, the factor $N^2$ in Eq.~(\ref{eq:spectrumfinalforw}) is now replaced by
\begin{equation}
\Bigl| \sum_{n=1}^{N} \eu^{\uimm\,\Delta k\,\hat{\boldsymbol{e}}_y\cdot\boldsymbol{r}_n }\Bigr|^2\approx N^2 \sinc^2{\left(\frac{\Delta k L }{2}\right)}.
\end{equation}
As done in Eq.~(\ref{eq:eta}), the sum was approximated with an integral over the length $L$, having assumed a constant linear density $N/L$. For the densities assumed here, ${\Delta k L }/{2}\approx 10^{-8}$, such that the factor $\sinc^2{\left({\Delta k L }/{2}\right)}$ is so close to 1 that it can be completely neglected \cite{PhysRevA.78.043409, JPSJ.30.518, arXiv:1203.4127}, as we did in Eqs.~(\ref{eq:spectrumintermediate}) and (\ref{eq:spectrumfinalforw}). Its presence, however, might have to be taken into account if a different experimental setting were considered.

\end{document}